\pdfoutput=1
\documentclass[]{smule}
\usepackage{libertinus}
\usepackage{amsmath}
\usepackage{amssymb}

\usepackage{amsmath,amsfonts,bm}

\def\eqref#1{equation~\ref{#1}}
\def\Eqref#1{Equation~\ref{#1}}

\def\1{\bm{1}}

\DeclareMathAlphabet{\mathsfit}{\encodingdefault}{\sfdefault}{m}{sl}
\SetMathAlphabet{\mathsfit}{bold}{\encodingdefault}{\sfdefault}{bx}{n}

\usepackage{url}
\usepackage{longtable}
\usepackage{wrapfig}

\title{TAPE: Learning General Audio Effect Representations}

\author[2*]{Phillip Long}
\author[1*]{Yongyi Zang}
\author[1]{Qihui Yang}
\author[1]{Christodoulos Benetatos}
\author[1]{Junfei Liu}
\author[1]{Randal Leistikow}
\author[2]{Julian McAuley}
\author[1]{Perry R. Cook}

\affiliation[1]{Smule Labs}
\affiliation[2]{University of California, San Diego}

\contribution[*]{These authors contributed equally.}

\abstract{
Musicians and mixing engineers often can decipher the signal processing chain they hear on a reference recording, then adapt it to new material.
We present TAPE (Transformations of Audio Parameter Embeddings), which estimates this processing from the processed reference audio alone.
An encoder trained to identify processors in eight major effect families maps the reference to an embedding.
A differentiable digital signal processing (DSP) chain decodes the embedding into editable parameters in one forward pass and applies them to a different unprocessed recording.
Its modules are equalization, compression, waveshaping, a small residual network, and reverberation.
TAPE then drives this change to shape the signal processing chain.
TAPE achieves lower spectral error than DeepAFx-ST, the previous state of the art for single-step regression, and lower spectral error than per-example search over a chain approximating ST-ITO's, while running 275$\times$ faster than that search. Because the decoder outputs DSP parameters, its chain can also run causally: a C++ implementation with the trained weights processed 128-sample buffers 1.8$\times$ faster than real time, with no missed deadlines.
}

\def\projectpageurl{https://pnlong.github.io/audio-effect-transfer/}
\def\projectpage{\url{\projectpageurl}}

\begin{document}

\maketitle

\section{Introduction}
\label{sec:introduction}

Engineers and musicians shape the timbre, dynamics, and space of a recording with audio effects such as equalization, compression, and reverberation~\citep{zolzer2002dafx,wilmering2020history}.
They also copy a treatment they have heard, such as the compression on a vocal, onto new material.
We aim to estimate that treatment from the processed reference alone.
The model should expose controls that a person knows how to use and run at the buffer sizes a performer plays through.
This is hard because professional processors are heterogeneous, their internals are rarely documented, and we observe them only through the audio they produce.

Most neural models of audio effects emulate one device or one effect type, in some cases so closely that listeners cannot tell the model from the device~\citep{wright2020realtime}.
Style-transfer systems instead extract an embedding from a processed reference and use it to drive a renderer.
DeepAFx-ST~\citep{steinmetz2022deepafx} predicts the controls of a differentiable equalizer and compressor in one forward pass, and DeepAFx-ST+~\citep{steinmetz2024st} adds reverberation and distortion.
ST-ITO~\citep{steinmetz2024st} covers many effect families by search.
For each reference, a gradient-free optimizer tunes a plugin chain until its output matches the reference in a learned effect embedding, rendering up to 1{,}600 candidates per example.
In these systems, one-pass prediction gives readable controls for a few processors, and ST-ITO needs a search for each example to cover many families.
Concurrent systems that predict whole chains are discussed in Section~\ref{sec:related-work}.

\begin{figure}[t]
\begin{center}
\includegraphics[width=\linewidth]{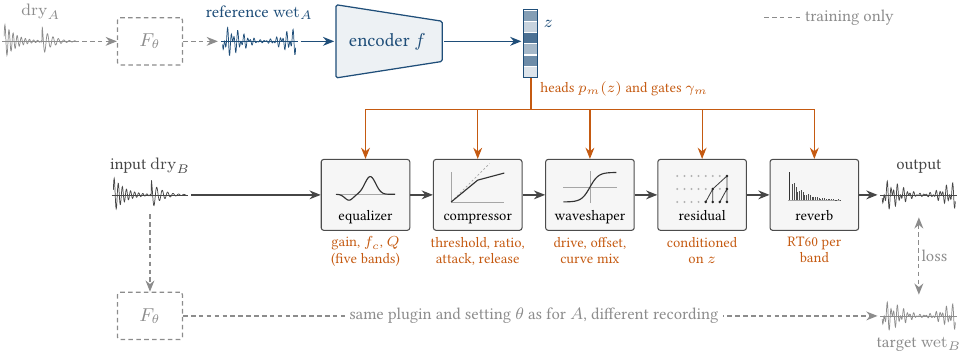}
\end{center}
\caption{Overview of TAPE.
The encoder $f$ maps a processed reference $\mathrm{wet}_{A}$ to an embedding $z$, and small heads map $z$ to the controls and gates of a five-module DSP chain that processes a different recording $\mathrm{dry}_{B}$.
In training (dashed), the same plugin configuration $\theta$ renders $A$ and $B$, and the output is compared with $\mathrm{wet}_{B}$.
Information about the source of $A$ in $z$ cannot help reconstruct $B$, but no loss removes it from $z$.}
\label{fig:system-overview}
\end{figure}

Our model, TAPE (Transformations of Audio Parameter Embeddings), represents audio-effect transformations in an embedding space.
An encoder $f$ maps a processed (wet) reference to an embedding $z$, and a decoder $g$ applies the treatment that $z$ describes to a new unprocessed (dry) recording (Figure~\ref{fig:system-overview}).
We train $f$ to identify 68 processors across speech, music stems, and instrument recordings, with the plugin as its only label.
We then train $g$ on pairs of recordings that share only the plugin, $g(\mathrm{dry}_B;z)\approx\mathrm{wet}_B$ with $z=f(\mathrm{wet}_A)$, so that cross-content reconstruction tests what $z$ captured.

The decoder is a factorized chain of differentiable DSP modules for equalization, compression, waveshaping, residual color, and reverberation.
Each module reads its parameters from $z$ through a small head, and the decoder outputs filter gains, compressor thresholds, reverberation times, and module gates that a person can inspect and edit.
A new reference needs one forward pass and no per-example search.
The modules also have causal approximations, which let the trained chain be streamed.
However, TAPE drives only its own chain, which approximates the 68 processors and has no modulation stage, and it cannot control plugins a user already owns.

Our contributions are as follows.
\begin{itemize}
  \item We train an effect encoder on 68 processors in eight families. Without family labels, its embeddings of the same family lie near one another in a t-SNE projection (Section~\ref{sec:representation}).
  \item We build a factorized differentiable DSP decoder that maps the embedding to the editable controls of a five-module chain (Section~\ref{sec:decoding}). After adversarial fine-tuning, every listener rated its transfer above that of its reconstruction-only counterpart, on two disjoint sets of pairs (Section~\ref{subsec:invertibility-results}).
  \item We compare TAPE with the released DeepAFx-ST models and with per-example search over ST-ITO's basic chain. TAPE had lower spectral error than either. On most pairs the search ended closer than TAPE to the reference in an effect embedding, but farther from the paired target (Section~\ref{subsec:search}).
  \item We adapt the trained chain to causal processing in C++. In simulated host callbacks on one CPU core, it streamed 128-sample buffers faster than real time (Section~\ref{subsec:realtime}).
\end{itemize}

\section{Related Work}
\label{sec:related-work}

\paragraph{Emulation of individual processors.}
Recurrent and convolutional networks can emulate a tube amplifier or a distortion circuit, in several cases in real time~\citep{damskgg2019deeplearningtube,wright2019real,damskagg2019real,wright2020realtime}.
Conditioning on the control settings lets one model cover the settings of a processor~\citep{hawley2019profiling}, and time-varying feature modulation extends black-box models to fuzz and compression~\citep{comunita2023modelling}.
This work continues a virtual-analog literature that includes white-box circuit simulation with wave digital filters and nonlinear state-space methods~\citep{fettweis1986wave,werner2016wave,yeh2008simulating,yeh2010automatedphysicalmodelin} and gray- and black-box system identification~\citep{eichas2016black,eichas2018gray,wright2022grey}.
\citet{pakarinen2009review} and \citet{vanhatalo2022review} survey the field, and \citet{comunita2025differentiable} compare black-box and gray-box architectures across many nonlinear devices.
Each of these models covers one device or a small set of devices.

\paragraph{Effect identification and parameter estimation.}
Classifiers that name the effect on a recording, or regress its settings, have been studied for guitar and studio effects~\citep{stein2010automatic,comunita2021guitar,hinrichs2022classification}.
Other work estimates the parameters of known processors from audio.
Examples include equalizer settings that transfer the frequency balance of an unpaired reference~\citep{mimilakis2020one}, the mixing parameters of a multitrack recording~\citep{colonel2021reverse}, and effect parameters learned by an autoencoder through differentiable effects~\citep{peladeau2024blind}.
Other systems recover the graph of processors behind a signal~\citep{lee2023blind} or an effect chain together with the dry signal~\citep{take2024audio}.
\citet{hinrichs2024blind} recover an approximate dry guitar signal and use it to train a neural copy of the reference effect.
RemFX~\citep{rice2023remfx} removes several effect classes by chaining removal models selected by a classifier.

\paragraph{Effect and style representations.}
DeepAFx-ST~\citep{steinmetz2022deepafx} encodes an input recording and a style reference with a shared encoder and predicts the parameters of a differentiable equalizer and compressor in one forward pass.
The model is trained without effect labels, and its embedding is evaluated with linear probes.
Contrastive encoders represent mixing effects in full mixes, rendered onto new content by a black-box network~\citep{koo2023music}, and in individual instruments within mixtures~\citep{yeh2025fxencoderpp}.
Diff-MST~\citep{vanka2024diff} predicts the controls of a differentiable mixing console from a reference mix.
\citet{chen2024zeroshot} condition a black-box amplifier generator on a contrastive tone embedding, including for amplifiers unseen in training.
ST-ITO~\citep{steinmetz2024st} is the closest prior work.
Its encoder, AFx-Rep, is pretrained on recordings from seven public datasets to predict which of 63 VST3 plugins and which preset turned an input into an output, from the embeddings of both.
ST-ITO then uses AFx-Rep as the objective of a CMA-ES search over the parameters of a user-supplied plugin chain, which may include non-differentiable plugins and plugins unseen in training.
A Gaussian prior over vocal-effect parameters fitted with DiffVox~\citep{yu2025diffvox} improves this search for vocals~\citep{yu2025gaussian}.
Our encoder, by contrast, sees only the processed signal, with plugin identity as its label, and our decoder predicts the controls directly, without search.

\paragraph{Concurrent work.}
LLM2Fx-Tools~\citep{doh2026llm2fxtools} has a language model emit an executable effect chain from a dry/wet pair through tool calls.
StemFX~\citep{cheng2026stemfx} autoregressively predicts a variable effect chain for each source-separated stem.
RelFx~\citep{liu2026beyond} learns a relative effect representation without a dry reference and transfers style by gradient search on a differentiable chain.
PEACE~\citep{braun2026peace} embeds Faust effect code and processed audio jointly to retrieve effect programs.
We do not compare with these systems.

\paragraph{Differentiable signal processing and language.}
DDSP~\citep{engel2020ddsp} trains classical oscillators and filters against reconstruction objectives while keeping controls a person can read.
Applications include differentiable IIR filters and equalizer matching~\citep{kuznetsov2020differentiable,nercessian2020neural} and automatic mixing with neural effect proxies~\citep{steinmetz2021automatic}.
Other work trains through non-differentiable plugins~\citep{martinez2021deepafx} or models the distribution of vocal effect settings~\citep{yu2025diffvox}.
\citet{hayes2024review} review differentiable DSP for music and speech synthesis.
Contrastive image--text and audio--text models align signals with language~\citep{radford2021clip,huang2022mulan,laionclap2023}.
Text2FX~\citep{chu2025text2fx} optimizes effect parameters toward a text description in CLAP space, and LLM2Fx~\citep{doh2025llm2fx} prompts a language model to predict effect parameters from a description.

\section{Formulation}
\label{sec:formulation}

Let $\mathrm{dry}$ be an unprocessed recording and $\theta$ a plugin configuration.
The plugin produces
\begin{equation}
  \mathrm{wet} = F_{\theta}(\mathrm{dry}),
  \label{eq:unknown-operator}
\end{equation}
and we observe only pairs $(\mathrm{dry},\mathrm{wet})$ and the identity of the plugin, never $F_{\theta}$ itself.
An encoder maps a wet clip to $z=f(\mathrm{wet})\in\mathbb{R}^{d}$, with $d=256$, and a decoder reconstructs the same treatment on a different dry recording,
\begin{equation}
  \widehat{\mathrm{wet}}_{B} = g(\mathrm{dry}_{B};\,z),\qquad z=f(\mathrm{wet}_{A}),
  \label{eq:invertibility}
\end{equation}
where $A$ and $B$ share $\theta$ but not waveform content (Figure~\ref{fig:system-overview}).
The decoder thus applies the processing estimated from one recording to another.
The encoder is trained to name the plugin on different kinds of content (Section~\ref{sec:representation}), and the decoder is then trained with $f$ frozen on pairs that share only $\theta$.
Neither objective enforces source invariance.

We constrain $g$ to a composition of differentiable DSP modules whose parameters are predicted from $z$,
\begin{equation}
  g(x;z) = \mathcal{M}_{K}(\,\cdots\,\mathcal{M}_{2}(\mathcal{M}_{1}(x;\,p_{1}(z));\,p_{2}(z))\,\cdots\,;\,p_{K}(z)),
  \label{eq:factorized-inverse}
\end{equation}
with $K=5$: equalization, compression, waveshaping, residual color, and reverberation.
Each $p_{m}(z)$ is a small multi-layer map to the actuators, or DSP parameters, of module $m$.
No module has a learned autoregressive loop, which makes a causal adaptation with a small per-buffer cost possible (Section~\ref{subsec:realtime}).
Plugin classification first shapes the embedding space $\mathcal{D}$ (Section~\ref{sec:representation}), the decoder is then fitted to the frozen encoder (Section~\ref{sec:decoding}), and an adversarial stage fine-tunes the system toward perceptual similarity (Section~\ref{subsec:joint-training}).
All audio is 48\,kHz mono, loudness-normalized to $-14$\,LUFS before processing, and analyzed in five-second windows.

\section{Learning the representation}
\label{sec:representation}

\begin{wrapfigure}{r}{0.46\linewidth}
\vspace{-1.2em}
\begin{center}
\includegraphics[width=\linewidth]{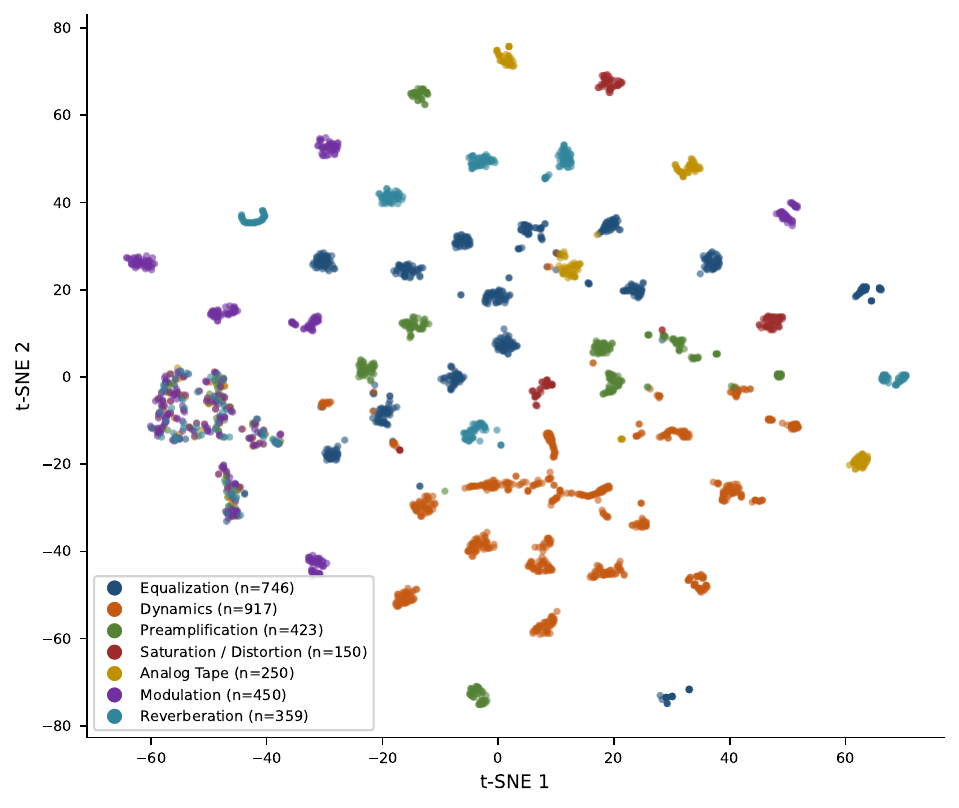}
\end{center}
\vspace{-0.8em}
\caption{t-SNE of held-out embeddings ($n=3{,}295$), colored by effect family.
The single miscellaneous plugin is not plotted.}
\label{fig:encoder-embedding-geometry}
\vspace{-1em}
\end{wrapfigure}

A log-mel front end at 48\,kHz (80 bands up to the Nyquist frequency) feeds a Conformer~\citep{gulati2020conformer}, and a learned query token pools its output into $z\in\mathbb{R}^{256}$ (Appendix~\ref{app:arch}).
We train $f$ with additive-margin softmax~\citep{wang2018amsoftmax} over the 68 plugin identities, and the variance and covariance terms of VICReg~\citep{bardes2022vicreg} keep the tight margins from collapsing or correlating the dimensions of $z$.
AFx-Rep~\citep{steinmetz2024st} is trained on a related task, but it also sees the unprocessed input and predicts the preset.
Plugin identity names neither the preset nor the dry source.
We chose this coarse label so that the encoder separates processors and leaves finer distinctions, such as settings, for the decoder to read from $z$.
The encoder is then frozen while the decoder is fitted, and the analyses below use it at that point.

\paragraph{Family structure.}
Figure~\ref{fig:encoder-embedding-geometry} projects held-out embeddings with t-SNE.
Processors of the same family form neighborhoods, although no loss named or grouped the families.
Since t-SNE preserves local neighborhoods but distorts global distances, the figure is evidence only that processors of the same family lie near each other in $\mathcal{D}$.

\paragraph{Plugin classification.}
The eight-layer encoder used in every decoding experiment reaches 72.5\% top-1 plugin accuracy on held-out wet clips, while a six-layer encoder reaches 78.7\% and deeper stacks and finer labels reduce accuracy (Appendix~\ref{app:encoder-ablations}).
With plugin-level supervision, configuration accuracy is near zero (Table~\ref{tab:encoder-ablations}).
Our reconstruction results do not show how much preset information the decoder obtains from $z$.

\section{A factorized, readable decoder}
\label{sec:decoding}

We build $g$ from DSP modules whose controls are physical quantities, such as filter gains and decay times, that a person can read and change.
A black-box generator conditioned on $z$ would expose no such controls, and it could hide whether the embedding captured the effect or the generator memorized a mapping.
Each module relaxes a classical operator so that it can be trained by gradient descent.
Appendix~\ref{app:operators} gives the operators in full and plots their responses.

\paragraph{Reading parameters from $z$.}
A small head maps $z$ to $d_{m}$ normalized actuators $p_{m}\in[0,1]^{d_{m}}$,
\begin{equation}
  p_{m,k} = \beta_{k}\,\sigma(v_{k}) + (1-\beta_{k})\,\tfrac{1}{2},\qquad
  [v_{k},\beta_{k}]=\mathrm{MLP}_{m}(z),
  \label{eq:ddsp-framing}
\end{equation}
where $\beta_{k}$ also passes through a sigmoid.
The gate $\beta_{k}$ blends each prediction toward the midpoint of its range, which stabilizes early training, and each $p_{m,k}$ is then mapped to a physical range (Table~\ref{tab:app-actuator}).
A module gate $\gamma_{m}$, a sigmoid of a small network on $z$, blends each stage with its input, $y=\gamma_{m}\,\mathcal{M}_{m}(x)+(1-\gamma_{m})\,x$, and $\gamma_{m}=0$ bypasses the stage.
In the later training stages, a weak penalty pulls the gates toward a binary pattern for each plugin category.

\paragraph{Equalization.}
Five peaking bands use the biquad response of \citet{bristow2016cookbook}, with gain, center frequency, and $Q$ predicted from $z$.
We apply the cascade as one multiplication on the FFT grid.
This approximates the IIR cascade by a circular convolution and avoids backpropagating through a long recurrence.

\paragraph{Compression.}
The hard knee of a feed-forward compressor~\citep{giannoulis2012digital} gives a predicted threshold no gradient while the compressor is not engaged.
We replace it with a softplus,
\begin{equation}
  G_{c}[n] = -\big(1-\tfrac{1}{r}\big)\log\!\big(1+e^{L[n]-T}\big),
  \label{eq:comp-gain}
\end{equation}
where $L[n]$ is a causal 10\,ms RMS level in decibels, $r$ the ratio, and $T$ the threshold.
The softplus matches the hard knee a few decibels from the threshold and stays differentiable in $T$ on both sides.
The gain is smoothed by one-pole attack and release filters, computed with prefix log-sums so that the recurrence can be differentiated.

\paragraph{Waveshaping.}
Choosing a waveshaping curve~\citep{brun1979digital,roads1979tutorial} is a discrete decision.
We run five classical curves in parallel (tanh saturation, hard clip, asymmetric tube, diode, and sine wavefolder) and mix them,
\begin{equation}
  y = \sum_{k=1}^{5} w_{k}\,s_{k}(a\,x+c),\qquad w=\mathrm{softmax}(\ell),\quad \ell\in[0,1]^{5},
  \label{eq:waveshaper}
\end{equation}
where $a$ is the predicted drive and $c$ a predicted asymmetry offset.
The logits $\ell$ are head outputs of \Eqref{eq:ddsp-framing}, so no curve receives more than $e/(e+4)\approx0.41$ of the weight.
A first-order tone filter and a makeup gain follow.

\paragraph{Residual color.}
The parametric stages cannot represent every coloration.
We append a causal temporal convolutional network~\citep{bai2018empirical} with FiLM conditioning on $z$~\citep{perez2018film}, trained after the parametric stages have converged.
The network has four dilated blocks with 32 channels and kernel size 13, a receptive field of 181 samples (3.8\,ms), and a module gate that can switch it off.
Among 23 residual architectures it ranked first on SI-SDR, with a grouped SRU and an S4D hybrid within 0.1\,dB (Appendix~\ref{app:nn-sweep}).
These differences are small next to the spread across pairs.
The comparison supports a small causal network but cannot rank block families.

\paragraph{Reverberation.}
Because feedback reverberators are hard to differentiate over tails that last seconds~\citep{valimaki2012fifty}, we synthesize the impulse response in closed form.
Band-filtered noise in six bands decays at a predicted reverberation time (RT60) per band, a broadband envelope damps the tail, and sparse early reflections and a direct impulse are added.
The three-second impulse is applied by FFT convolution.
A small ConvNeXt network conditioned on $z$ refines the STFT of this impulse, and the evaluated decoder blends the two through a learned weight.

\begin{figure}[t]
\begin{center}
\includegraphics[width=\linewidth]{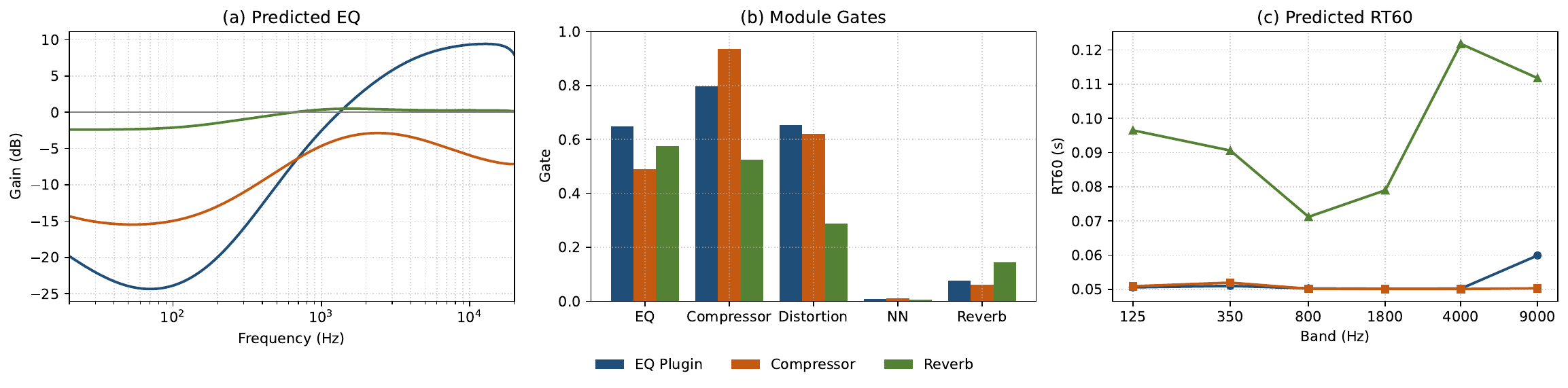}
\end{center}
\caption{Controls predicted from $z$ for three held-out embeddings (EQ Plugin, Compressor, Reverb).
(a)~Five-band EQ magnitude response.
(b)~Module gates (distortion is the waveshaper, NN the residual).
(c)~RT60 per band.}
\label{fig:predicted-controls}
\end{figure}

\paragraph{Reading the controls.}
Figure~\ref{fig:predicted-controls} shows the EQ curve, module gates, and RT60 profile that the trained decoder predicts for three held-out embeddings.
Each of the equalizer, compressor, and reverb gates is highest for its matching embedding.
The gates are graded: the equalizer, compressor, and distortion gates are partly open (0.29 to 0.94) for all three embeddings, and the residual gate is nearly closed.
The reverb gate is 0.14 for the reverb embedding and 0.06 to 0.08 for the others.
Only the reverb embedding's RT60 rises above the floor of about 50\,ms, to 0.07--0.12\,s.

\subsection{Training the decoder}
\label{subsec:joint-training}

The modules interact.
An equalizer boost that a compressor later tames produces conflicting gradients if each head is trained against reconstruction alone.
We compared three curricula on the same frozen $z$: heads trained separately, heads pretrained separately and then trained jointly, and heads trained jointly from scratch (Appendix~\ref{app:curricula}).
Joint training from scratch had the lowest spectral error.
However, listeners gave the two-stage curriculum the highest mean score (Study~2).
The decoder evaluated below starts from the joint-scratch checkpoint.
We have not tested the two-stage checkpoint as a starting point.

\paragraph{Adversarial fine-tuning.}
\label{subsec:gan-method}
After the heads and the residual converge, we fine-tune $g$ adversarially with feature matching~\citep{kong2020hifigan}, using critics that see multi-period waveforms, multi-resolution STFT magnitudes, frequency sub-bands, and multi-scale constant-Q transforms.
From the same checkpoint we trained two variants that differ only in the adversarial loss: a relativistic loss as in R3GAN~\citep{huang2024gan} and a least-squares loss~\citep{mao2017lsgan}, both with R1 and R2 gradient penalties.
Each experiment below reports results from one such fine-tuned checkpoint, noted where it differs across experiments.
This stage updates all parameters, the encoder included, at a learning rate of $5\times10^{-6}$.
The decoding experiments use the encoder of the fine-tuned checkpoint, which may differ from the one analyzed in Section~\ref{sec:representation}.
Before-and-after comparisons measure the whole stage, which also adds feature matching and continues training.

\section{Experimental setup}
\label{sec:experimental-setup}

\paragraph{Data.}
\label{subsec:dataset}
Training draws five-second wet/dry pairs from nearly two million mono renders of about 300 source recordings through randomized configurations of 68 plugins: 19 dynamics processors, 14 equalizers, 9 modulation effects, 9 preamplifiers, 8 reverberators, 5 analog-tape emulations, 3 saturation or distortion units, and 1 multi-tool (Appendix~\ref{app:plugins}).
Sources come from MUSDB18~\citep{rafii2019musdb18}, GuitarSet~\citep{xi2018guitarset}, Carnatic Varnam~\citep{koduri2014intonation}, Bach Violin~\citep{dong2022deep}, LibriSpeech~\citep{panayotov2015librispeech}, and synthetic test signals.
The encoder has to recognize a processor on speech, guitar stems, and instrument recordings that share nothing except the plugin.
Here ``dry'' means unprocessed by our plugins, since some sources, such as MUSDB18 stems, are already produced.
Sources are resampled to 48\,kHz and loudness-normalized, and each pair is a five-second window that starts at least one second into the render.
Tracks are split 80/10/10 into training, validation, and test sets by a hash of the track identifier, while plugin configurations are shared across splits.
For MUSDB18 the identifier names a stem, and stems of one song can fall into different splits.
The saved evaluation outputs do not record track identifiers.
The evaluation pairs are therefore held out only in the sense of this nominal split.
An effect-score heuristic redraws presets whose wet signal is close to the dry input (Appendix~\ref{app:effect-score}).

\paragraph{Metrics.}
\label{subsec:evaluation}
Our main metric is multi-resolution STFT L1~\citep{steinmetz2020auraloss}, the mean absolute difference between natural-log STFT magnitudes over FFT sizes 512, 1024, and 2048 (Hann window, hop of a quarter frame).
We also report time-domain L1, SI-SDR~\citep{leroux2019sdr}, log-spectral distance, and a CLAP distance, $1-\cos$ between the CLAP audio embeddings~\citep{laionclap2023} (HTSAT, unfused checkpoint) of output and target.
These metrics rank systems differently from one another and from listeners, especially on reverberation and subtle color.
We treat listening as the primary test.

\paragraph{Listening tests.}
Four multi-stimulus panels in the MUSHRA format~\citep{bs1534,schoeffler2018webmushra}, each with 20 trials, ran in one headphone session with overlapping listeners, who rated the similarity of each stimulus to a reference from 0 to 100, with a hidden reference.
Without the standard low-pass anchors, we read the scores as within-trial similarity ratings and treat the unprocessed input, where it appears, as a control condition.
Stimuli were 192\,kb/s MP3, and one gain per trial placed the loudest peak at $-3$\,dBFS.
Study~1 compares the adversarially fine-tuned decoder with its reconstruction-only counterpart and with unprocessed audio, and Study~3 repeats the first contrast on disjoint pairs.
Both use cross-content held-out pairs.
Study~2 compares the three curricula and Study~4 the two reverberation impulses, on same-content validation clips (Appendices~\ref{app:curricula} and~\ref{app:reverb-tiers}).
After we dropped one listener who failed the reference check, Studies~1--4 had eight, six, seven, and nine listeners.
The audio is on the project page.\footnote{\projectpage}

\section{Experiments}
\label{sec:results}

\subsection{Cross-content transfer}
\label{subsec:invertibility-results}

We extract $z$ from track $A$ and ask the decoder (R3GAN variant) to reproduce the treatment on track $B$.
The 60 held-out pairs were drawn with a fixed seed from test-split tracks, and each contains two tracks rendered through the same configuration.
The adversarial stage lowered the mean STFT L1 from 2.071 to 1.120, time-domain L1 from 0.200 to 0.075, and CLAP distance from 0.464 to 0.220, and it raised SI-SDR from $-29.38$ to $-16.80$\,dB.
Among families with at least six pairs, dynamics had the lowest STFT L1 (0.927) and equalization the highest (1.299), and chains of several processors were close to the mean (1.176; Appendix~\ref{app:transfer}).

\begin{table}[t]
\caption{Listening results for cross-content transfer (similarity to the reference, 0--100).
Means and SDs are over all ratings, and 95\% CIs are bootstrap intervals over per-listener means.
Scores are for the least-squares variant.
Study~1 includes the unprocessed input as a control; Study~3 repeats the main contrast on new pairs.}
\label{tab:mushra}
\begin{center}
\begin{small}
\begin{tabular}{lccclcc}
\toprule
 & \multicolumn{3}{c}{\textbf{Study 1} ($N=8$)} & & \multicolumn{2}{c}{\textbf{Study 3} ($N=7$)} \\
\cmidrule(lr){2-4}\cmidrule(lr){6-7}
\textbf{Condition} & \textbf{Mean} & \textbf{SD} & \textbf{95\% CI} & & \textbf{Mean} & \textbf{SD} \\
\midrule
Adversarially fine-tuned & \textbf{72.5} & 23.0 & [66.3, 78.5] & & \textbf{72.4} & 27.3 \\
Unprocessed      & 47.6 & 27.0 & [38.8, 56.4] & & -- & -- \\
Reconstruction only          & 41.3 & 25.0 & [33.0, 50.0] & & 38.4 & 24.6 \\
\bottomrule
\end{tabular}
\end{small}
\end{center}
\end{table}

In Study~1, every listener's mean score was higher for the fine-tuned decoder than for the reconstruction-only decoder and for unprocessed audio (Table~\ref{tab:mushra}).
Both contrasts gave $p=0.0078$ in exact two-sided Wilcoxon signed-rank tests on per-listener means, or $0.0156$ after Holm correction.
Study~3 repeated the first contrast on disjoint pairs, in the same session and with an overlapping panel, and the gain held for every listener ($p=0.0156$).
Listeners rated the reconstruction-only decoder below the unprocessed input, although on the 60 pairs above its STFT L1 (2.071) was lower than the input's (2.335).
Since these tests treat listeners as the unit and do not resample stimuli, they do not account for item variation.

\subsection{Comparison with per-example search and with DeepAFx-ST}
\label{subsec:search}

Per-example search in the manner of ST-ITO~\citep{steinmetz2024st} matches the reference in an effect embedding.
We test whether it brings outputs closer to the target than one forward pass of TAPE.
We also run the released DeepAFx-ST models~\citep{steinmetz2022deepafx} on the same 60 pairs.
Every method sees only $\mathrm{wet}_{A}$ and $\mathrm{dry}_{B}$.
The withheld $\mathrm{wet}_{B}$ is used only for scoring, except by one oracle.

\paragraph{Search.}
ST-ITO's main chain consists of five third-party VST3 plugins that are not distributed with its public code.
Our search instead controls the basic chain of ST-ITO's public code: an equalizer, a compressor, a distortion, a delay, and a reverberator.
These are the same five effect types as the pedalboard chain of the original paper.
CMA-ES minimizes the cosine distance to $\mathrm{wet}_{A}$, either in AFx-Rep or in our encoder $f$, and random search with the AFx-Rep objective serves as a control.
The \emph{full} setting follows ST-ITO's public evaluation script, with 32 generations of 128 candidates ($\sigma=0.33$) started at the midpoint of every range, for 4{,}097 renders per example with early stopping disabled.
This exceeds the budget described in the ST-ITO paper, a population of 64 for at most 25 steps (1{,}600 renders) with early stopping.
A \emph{reduced} setting with 128 renders also differs in population, step size, initialization, and padding.
The two settings thus do not form a budget ablation (Appendix~\ref{app:search-family}).
\emph{TAPE, then chain search} applies the same search to the chain placed after TAPE's output, and keeps TAPE's output if no candidate improves the objective.
We also report STFT L1 after peak-normalizing output and target.
The chain's input and output are peak-normalized as in ST-ITO, while TAPE preserves level.
The comparison does not rank TAPE and ST-ITO.
Besides differing from ST-ITO's VST chain, the basic chain differs from TAPE's: it has a delay and no residual.
ST-ITO was also designed for style transfer judged by listeners.

\paragraph{DeepAFx-ST.}
We ran the two released models, trained on speech and on music, with their official recipe.
Input and reference are resampled to 24\,kHz and peak-normalized to $-12$\,dBFS, and the output is peak-normalized and resampled to 48\,kHz.
Since a full-band error heavily penalizes an output with no content above 12\,kHz, we also score every method after resampling output and target to 24\,kHz and peak-normalizing each.
As an ablation, and as DeepAFx-ST's authors did for 44.1\,kHz audio, we also render the predicted controls on the 48\,kHz input.
The DeepAFx-ST renderer contains only an equalizer and a compressor, and the released models were trained on other recordings.
This comparison tests the released models outside their training distribution.
It does not test the approach trained on our pairs.

\paragraph{Measurement.}
All runs use one seed, the same GPU, and eight CPU threads.
Times include encoding, rendering, and optimization.
TAPE runs with PyTorch's default TF32 convolutions; disabling TF32 leaves the weights unchanged and raises its mean STFT L1 from 1.120 to 1.177.
Confidence intervals come from 2{,}000 bootstrap resamples over effect configurations, and $p$-values from paired Wilcoxon signed-rank tests over the 60 pairs.

\begin{table}[t]
\caption{Search at the full setting and the released DeepAFx-ST models (60 held-out pairs, means over pairs).
STFT L1 is measured against the withheld $\mathrm{wet}_{B}$ at 48\,kHz.
\emph{Peak} normalizes output and target to unit peak at 48\,kHz, and \emph{Peak, 24\,kHz} first resamples both to 24\,kHz and then normalizes each to unit peak.
\emph{AFx} is the cosine distance to $\mathrm{wet}_{A}$ in AFx-Rep (not computed for DeepAFx-ST or with TF32 disabled).
\emph{Wins} counts pairs on which a method has a lower STFT L1 (48\,kHz, unnormalized) than TAPE.
Renders count evaluations of a chain per example; the corrective searches add TAPE's pass and skip the final render when they keep TAPE's output.
Time is the median wall-clock seconds per five-second clip.
${}^{\ddagger}$Controls predicted at 24\,kHz, rendered on the 48\,kHz input.
Appendix~\ref{app:search-family} gives the reduced setting and searches in TAPE's own space.}
\label{tab:fullbudget}
\begin{center}
\begin{footnotesize}
\setlength{\tabcolsep}{4pt}
\begin{tabular}{llrrccccc}
\toprule
\textbf{Method} & \textbf{Objective} & \textbf{Renders} & \textbf{Time} & \textbf{STFT L1} & \textbf{Peak} & \textbf{Peak, 24\,kHz} & \textbf{AFx} & \textbf{Wins} \\
\midrule
Unprocessed $\mathrm{dry}_{B}$ & -- & 0 & -- & 2.335 & 1.808 & 1.724 & 0.894 & 5 \\
TAPE, one pass & -- & 1 & 0.52 & 1.120 & 1.391 & 1.310 & 0.660 & -- \\
TAPE, one pass, TF32 disabled & -- & 1 & 0.53 & 1.177 & 1.448 & 1.323 & -- & -- \\
\midrule
\multicolumn{9}{l}{\textit{Search over ST-ITO's basic chain, full setting}} \\
ST-ITO search & AFx-Rep & 4{,}097 & 143 & 2.906 & 2.317 & 2.150 & 0.259 & 2 \\
ST-ITO search & $f$ & 4{,}097 & 205 & 2.498 & 1.703 & 1.580 & 0.652 & 5 \\
Random search & AFx-Rep & 4{,}097 & 142 & 2.936 & 2.175 & 2.005 & 0.373 & 2 \\
TAPE, then chain search & AFx-Rep & 4{,}098 & 144 & 2.701 & 2.154 & 2.079 & 0.250 & 4 \\
TAPE, then chain search & $f$ & 4{,}098 & 205 & 2.454 & 1.769 & 1.684 & 0.642 & 3 \\
\midrule
\multicolumn{9}{l}{\textit{Released DeepAFx-ST models (equalizer and compressor)}} \\
DeepAFx-ST, speech & -- & 1 & 0.30 & 5.511 & 6.286 & 2.498 & -- & 0 \\
DeepAFx-ST, music & -- & 1 & 0.29 & 5.226 & 5.884 & 2.097 & -- & 0 \\
DeepAFx-ST, speech, 48\,kHz render${}^{\ddagger}$ & -- & 1 & 0.20 & 2.548 & 2.538 & 2.347 & -- & 0 \\
DeepAFx-ST, music, 48\,kHz render${}^{\ddagger}$ & -- & 1 & 0.23 & 2.228 & 2.055 & 1.884 & -- & 7 \\
\bottomrule
\end{tabular}
\end{footnotesize}
\end{center}
\end{table}

\begin{figure}[t]
\begin{center}
\includegraphics[width=0.86\linewidth]{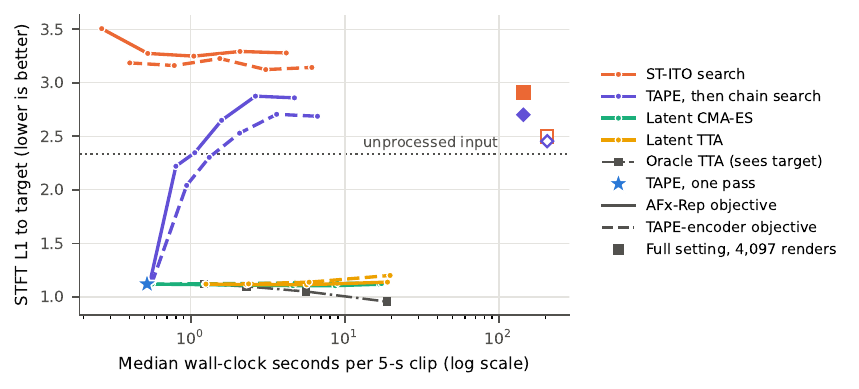}
\end{center}
\caption{Reconstruction error against wall-clock cost (60 pairs).
Lines follow each search as its budget grows at the reduced setting: 8 to 128 renders of the basic chain, 8 to 32 latent candidates, or 1 to 16 TTA steps.
Solid lines optimize AFx-Rep and dashed lines optimize $f$.
Squares (ST-ITO search) and diamonds (TAPE, then chain search) mark the full setting of Table~\ref{tab:fullbudget}, filled for AFx-Rep and open for $f$.
The star is TAPE's single forward pass, and the dotted line is the unprocessed input.}
\label{fig:search}
\end{figure}

\paragraph{Search results.}
At both settings, no method available at inference had a significantly lower mean error than TAPE's forward pass.
At the full setting, the AFx-Rep search had a mean STFT L1 of 2.91, above both TAPE's 1.12 and the input's 2.34 (Table~\ref{tab:fullbudget}), a difference from TAPE of $+1.79$ [$+1.37$, $+2.30$] ($p<10^{-10}$).
The search had lower error than TAPE on 2 of the 60 pairs and took 143\,s per clip, about 275 times TAPE's time.
With $f$ as the objective, the search reached 2.50 ($+1.38$ [$+1.09$, $+1.69$] against TAPE), or 1.70 after peak normalization, against 1.81 for the input and 1.39 for TAPE.
Starting the search from TAPE's output moved it away from the target (2.70 with AFx-Rep and 2.45 with $f$).
Every search remained above TAPE at the common 24\,kHz bandwidth ($p<0.003$).
At the reduced setting, every search over the basic chain had a higher mean error than both TAPE and the input.
The ST-ITO search had a higher error than TAPE in every effect family, at either setting (Appendix~\ref{app:search-family}).
The gap remains after peak normalization, although peak-normalizing the chain's input, as ST-ITO does, also changes how its compressor and distortion act.

\paragraph{Embedding distance and paired error.}
The searches did lower the distance they optimized.
From the reduced to the full setting, the AFx-Rep search lowered its AFx-Rep distance from 0.39 to 0.26, but its STFT L1 did not fall significantly (from 3.28 to 2.91; $p=0.08$).
On 56 of the 60 pairs it moved closer to the reference in AFx-Rep than TAPE did while moving farther from the target (49 at the reduced setting).
At both settings, CMA-ES reached a lower AFx-Rep distance than random search, but its error was not significantly lower (at the full setting, 0.26 versus 0.37, and STFT L1 2.91 versus 2.94; $p=0.68$).
Within each pair, we rank-correlated the AFx-Rep distance to $\mathrm{wet}_{A}$ with the error to $\mathrm{wet}_{B}$ over the outputs of TAPE and the nine reduced-setting searches and adaptations in Table~\ref{tab:search} (Appendix~\ref{app:search-family}).
The median Spearman correlation across pairs was $-0.17$, against 0.65 for distance in $f$, although the outputs come from two different renderers.
In these evaluations, minimizing effect-embedding distance did not reliably improve paired reconstruction.

Searching in TAPE's own space did not help either.
Latent CMA-ES, which perturbs $z$ around $f(\mathrm{wet}_{A})$ and decodes each candidate with $g$, changed the mean STFT L1 by less than 0.005 under either objective ($p\ge0.33$).
Sixteen Adam steps on $z$ (latent TTA) gave no significant change with AFx-Rep ($+0.018$, $p=0.86$) and raised the error by 7\% with $f$ ($p=0.02$, uncorrected).
An oracle that took the same steps against a multi-resolution STFT loss to $\mathrm{wet}_{B}$ lowered the error by 15\% and improved 55 of the 60 pairs.
The decoder can thus render outputs closer to the target than the one that $f(\mathrm{wet}_{A})$ selects.

\paragraph{DeepAFx-ST results.}
The released DeepAFx-ST models had the highest error of all methods at 48\,kHz (5.51 for the speech model and 5.23 for the music model), mostly due to the missing content above 12\,kHz.
At the common 24\,kHz bandwidth with peak normalization, the music model reached 2.10, against 1.31 for TAPE and 1.72 for the unprocessed input, and TAPE had the lower error on 55 of the 60 pairs ($p<10^{-8}$).
With its controls rendered on the 48\,kHz input, the music model reached 1.88 at this bandwidth, still above the input's error.
The gap also held within the dynamics and equalization families (20 pairs), where the music model reached 2.08, against 1.35 for TAPE and 1.86 for the input.
The released models took about 0.3\,s per clip, against 0.52\,s for TAPE.

\subsection{Streaming the decoder}
\label{subsec:realtime}

A musician can play through an effect only if it adds little latency~\citep{lago2004latency}.
The encoder and the control heads run once per reference, and only the chain of $g$ has to run per buffer.
We adapted the chain to causal processing in C++ and carry all filter, envelope, and convolution state across buffers.
The chain adds no buffering or lookahead beyond the host buffer, although its filters and reverberator keep their own phase delay.
The offline equalizer, tone filter, and mean removal have no exact causal counterpart.
The FFT-grid equalizer becomes a cascade of five biquads, the zero-phase tone filter becomes a first-order low-pass with the same cutoff, and clip-mean removal becomes a 20\,Hz DC blocker.
The streamed chain is thus a causal variant of $g$ and not a sample-exact port.
Appendix~\ref{app:benchmarks} measures each substitution.

\begin{table}[t]
\caption{Streaming the trained chain on one CPU core at 48\,kHz, over the 60 five-second dry inputs of Section~\ref{subsec:search}, in simulated host callbacks.
Time is the mean processing time per clip, excluding the once-per-reference conditioning, and RTF is the clip duration divided by that time.
p99 is the 99th percentile of callback time, averaged over clips; Worst is the slowest callback over all clips; Missed counts callbacks that exceeded the budget.}
\label{tab:benchmarks-main}
\begin{center}
\begin{small}
\begin{tabular}{rrrcrrr}
\toprule
\textbf{Buffer} & \textbf{Budget (ms)} & \textbf{Time (s)} & \textbf{RTF} & \textbf{p99 (ms)} & \textbf{Worst (ms)} & \textbf{Missed} \\
\midrule
64   & 1.33  & 3.32 & 1.50 & 0.92  & 1.76  & 21 of 225{,}000 \\
128  & 2.67  & 2.84 & 1.76 & 1.56  & 2.38  & 0 of 112{,}500 \\
256  & 5.33  & 2.61 & 1.92 & 2.85  & 3.64  & 0 of 56{,}280 \\
512  & 10.67 & 2.52 & 1.99 & 5.47  & 6.28  & 0 of 28{,}140 \\
1{,}024 & 21.33 & 2.48 & 2.02 & 10.72 & 11.77 & 0 of 14{,}100 \\
\bottomrule
\end{tabular}
\end{small}
\end{center}
\end{table}

\paragraph{Speed.}
We streamed the chain of the R3GAN variant with its trained weights on one CPU core and simulated host callbacks offline (Table~\ref{tab:benchmarks-main}).
Conditioning, that is, the encoder, the control heads, and the impulse generator, also runs in C++ and takes 1.43\,s per reference.
With 128-sample buffers (2.67\,ms), a callback took 1.51\,ms on average, 1.76 times faster than real time, and none of the 112{,}500 callbacks exceeded its budget.
With 64-sample buffers the chain ran 1.50 times faster than real time on average.
However, 21 of 225{,}000 callbacks, in 20 of the 60 clips, missed their deadline.
The program allocates memory inside callbacks, its timing was not isolated from other workloads, and the budgets exclude the buffering of an audio interface.
These measurements do not establish a hard real-time guarantee.
Appendix~\ref{app:benchmarks} describes a second, allocation-free implementation of the same chain, which we have run only with random weights.

\paragraph{Fidelity.}
On every clip, the outputs streamed at different buffer sizes agree at a signal-to-error ratio above 111\,dB.
In a causality check on one clip, replacing the input from a point inside a callback onward changed the earlier output by at most $4.5\times10^{-8}$.
Run on whole clips with the offline operators, the C++ program matches PyTorch in strict single precision (TF32 disabled) at a waveform signal-to-error ratio of 67.9\,dB on average and at least 34.5\,dB.
The streamed output's error against the target is close to the offline error at this precision, with a mean STFT L1 of 1.179 against 1.177 (difference $+0.002$ [$-0.009$, $+0.016$]; $p=0.12$) and of 1.506 against 1.448 after peak normalization ($p=0.17$).
At 24\,kHz with peak normalization the streamed output had a higher error (1.366 against 1.323; $p=0.04$, uncorrected).
The streamed and offline waveforms differ more, at a signal-to-error ratio of 22.7\,dB on average and from $-16.6$ to $49.8$\,dB across clips.
Resetting the state at every 256-sample buffer instead raised STFT L1 by 1.55 [1.30, 1.81].
We have not tested whether listeners hear the difference between streamed and offline outputs.

\section{Discussion}
\label{sec:discussion}

\paragraph{Evaluation.}
Objective metrics and listeners disagreed in our studies: joint training from scratch had the lowest STFT L1 but scored about the same as dry audio (Study~2).
Refining the reverberation impulse halved its time-domain L1 but worsened its STFT L1, without a detected listening difference.
With six to nine overlapping listeners per panel and no low anchor, we rely on within-listener contrasts and on the repeat of the main contrast in Study~3.
Claims about absolute quality would need a larger panel and item-level modeling.
The objective and listening results also come from two variants of the adversarial stage, and Studies~2 and~4 and the ablations use same-content validation clips.

\paragraph{Inference-time objectives.}
In Section~\ref{subsec:search}, search over ST-ITO's basic chain and gradient steps on $z$ both lowered distance in an effect embedding.
Neither lowered the average error against the target, but an oracle loss against the target did.
The oracle lowers the error by changing only $z$, which places part of the remaining error in the choice of $z$.
The chain mismatch noted in Section~\ref{subsec:search} also means that our search does not separate the search method from the renderer.
Running ST-ITO on its VST chain and searching TAPE's own actuators from a neutral start would do so.
Retraining DeepAFx-ST or DeepAFx-ST+ on our pairs would remove the mismatch in training data but not in renderers.
ST-ITO can drive unseen and non-differentiable plugins, and TAPE cannot.

\paragraph{Scope and future work.}
TAPE covers 68 mono processors.
The chain has no stage driven by a low-frequency oscillator, and the residual's 3.8\,ms receptive field cannot realize one.
The nine modulation plugins and tape wow and flutter are only approximated.
Unseen plugins and stereo processing would need new rendering, a stereo encoder input, and a stereo chain.
Comparing embeddings of different recordings processed by the same plugin would measure source invariance directly.
References that share the plugin but not the setting would show how much of the transfer relies on setting-specific information in $z$.
An unconstrained decoder on the same frozen $z$ would show what the readable chain costs in fidelity.
Other next steps are refining the joint-finetune checkpoint, a reconstruction-only continuation with the same budget and trainable parameters as the adversarial stage, and training $g$ with its causal operators.
For streaming, the trained program still has to run without allocation in an audio-device callback.

\section{Conclusion}
\label{sec:conclusion}

We presented TAPE, which embeds the processing on a reference recording with an encoder trained to identify 68 processors.
A factorized differentiable DSP chain decodes the embedding into editable controls in one forward pass.
Before fine-tuning, the encoder places processors of the same family near each other in a t-SNE projection, without family labels.
After a least-squares adversarial stage, listeners rated the decoder's transfer above its reconstruction-only counterpart on two disjoint sets of pairs, and above the unprocessed input in Study~1.
The variant fine-tuned with the R3GAN loss had lower spectral error than the released DeepAFx-ST models and than search over ST-ITO's basic chain at 4{,}097 renders per example, in about 1/275 of the search's time.
On most pairs, the search ended closer to the reference in an effect embedding than TAPE did but farther from the target.
Better inference-time objectives remain an open problem.
In simulated host callbacks, a causal C++ version of the trained chain ran faster than real time on average on one CPU core, and its 48\,kHz spectral error was close to that of the offline decoder.
TAPE drives only its own chain and cannot control plugins a user already owns.
The chain still has no modulation stage, and stereo processing and unseen plugins remain future work.

\bibliographystyle{plainnat}
\bibliography{main}

\newpage
\beginappendix

\section{Encoder ablations}
\label{app:encoder-ablations}

\begin{table}[h]
\caption{Encoder depth and training objective.}
\label{tab:encoder-ablations}
\begin{center}
\begin{small}
\begin{tabular}{llccc}
\toprule
\textbf{Depth} & \textbf{Objective} & \textbf{Plugin acc.\ (\%)} & \textbf{Config acc.\ (\%)} & \textbf{Val.\ loss} \\
\midrule
6 & Plugin AM-Softmax & 78.7 & 0.01 & 0.875 \\
8 & Plugin AM-Softmax & 72.5 & 0.00 & 1.115 \\
6 & Config AM-Softmax & 63.9 & 39.0 & 3.908 \\
8 & Config AM-Softmax & 55.9 & 25.6 & 4.341 \\
6 & Hierarchical & 44.3 & 0.89 & 5.074 \\
8 & Hierarchical & 46.9 & 0.57 & 5.237 \\
10 & Plugin AM-Softmax & 46.6 & 0.00 & 2.001 \\
12 & Plugin AM-Softmax & 54.8 & 0.01 & 1.758 \\
\bottomrule
\end{tabular}
\end{small}
\end{center}
\end{table}

Plugin-level angular margins give the highest plugin accuracy in this grid, and configuration-level margins buy some preset accuracy at a much higher validation loss.
The eight-layer encoder used for all decoding experiments is the second most accurate classifier, after the six-layer one.

\section{Decoder operators}
\label{app:operators}

\begin{figure}[h]
\begin{center}
\includegraphics[width=\linewidth]{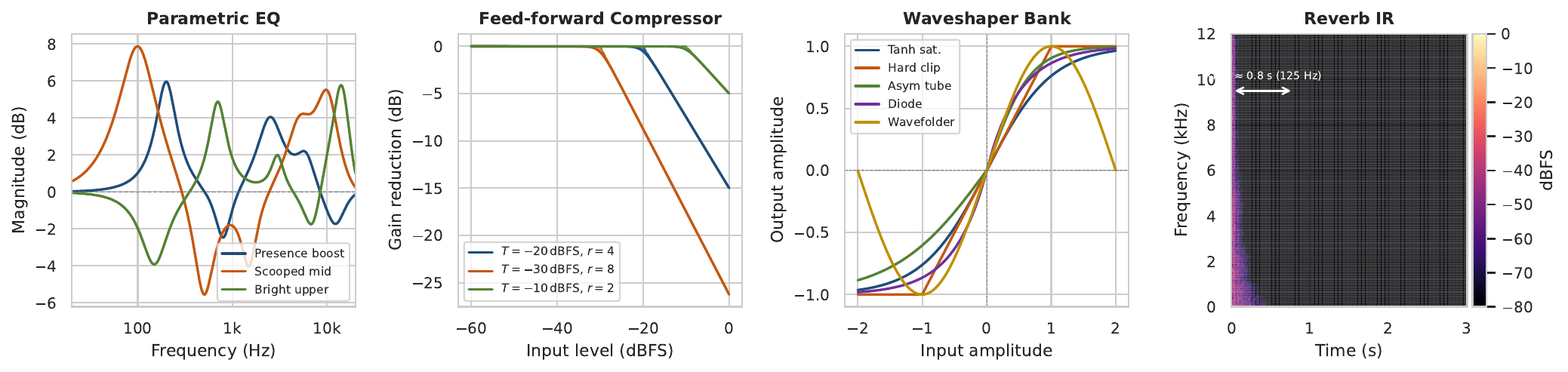}
\end{center}
\caption{Module responses of the decoder.
From left to right: five-band peaking EQ; compressor gain reduction (small shaded band: softplus versus hard knee); five parallel waveshapers; log-magnitude spectrogram of a parametric reverberation impulse.}
\label{fig:dsp-module-responses}
\end{figure}

This appendix gives the operators of Section~\ref{sec:decoding} in full.
Figure~\ref{fig:dsp-module-responses} shows their responses, and Table~\ref{tab:app-actuator} gives the actuator ranges.

\paragraph{Gates.}
In the residual and adversarial stages (Appendix~\ref{app:hparams}), a mean-squared penalty pulls the gates toward a binary activation pattern defined for each plugin category.
For example, an equalizer should not open the reverberator.
The penalty's weight is small enough that the reconstruction loss can override it, so the gates are a readout of what $z$ captured (Figure~\ref{fig:predicted-controls}b).

\paragraph{Equalization.}
A parametric peaking equalizer is classically a cascade of Bristow--Johnson biquads~\citep{bristow2016cookbook}.
Band $k$ with gain $G_{k}$\,dB, center frequency $f_{k}$, and quality $Q_{k}$ has
\begin{align}
  A_{k} &= 10^{G_{k}/40},\quad
  \omega_{k} = \frac{2\pi f_{k}}{f_{s}},\quad
  \alpha_{k} = \frac{\sin\omega_{k}}{2Q_{k}}, \notag \\
  H_{k}(z) &= \frac{1 + \alpha_{k} A_{k} - 2\cos\omega_{k}\,z^{-1} + (1-\alpha_{k} A_{k})\,z^{-2}}
                  {1 + \alpha_{k}/A_{k} - 2\cos\omega_{k}\,z^{-1} + (1-\alpha_{k}/A_{k})\,z^{-2}}.
  \label{eq:eq-biquad}
\end{align}
The time-domain recurrence is differentiable in the coefficients.
Our decoder uses five peaking bands, with $(f_{k},G_{k},Q_{k})$ predicted from $z$, and applies the cascade as a single frequency-domain multiplication,
\begin{equation*}
  y=\mathrm{iRFFT}\big(\mathrm{RFFT}(x)\textstyle\prod_{k=1}^{5}H_{k}\big).
\end{equation*}
This evaluates the response of the IIR cascade on the FFT grid, which amounts to a circular convolution with a time-aliased impulse response, and avoids backpropagating through a long recurrence.

\paragraph{Compression.}
A feed-forward compressor~\citep{giannoulis2012digital} measures a level $L[n]$, applies a hard-knee gain computer
\begin{equation}
  G_{c}^{\mathrm{hard}}[n] =
  \begin{cases}
    -\big(1-\tfrac{1}{r}\big)\big(L[n]-T\big), & L[n]>T, \\
    0, & \text{otherwise},
  \end{cases}
  \label{eq:comp-gain-hard}
\end{equation}
and smooths $G_{c}$ with one-pole attack and release.
The kink at $L[n]=T$ has zero gradient on the unengaged side, so a predicted threshold receives no learning signal while the compressor is off.
Level detection is a causal 10\,ms RMS,
\begin{equation}
  L[n] = 10\log_{10}\!\left(\frac{1}{W}\sum_{m=0}^{W-1}x^{2}[n-m]\right),
  \label{eq:comp-level}
\end{equation}
and the softplus of \Eqref{eq:comp-gain} replaces the hard knee of \Eqref{eq:comp-gain-hard}.
Attack and release coefficients $\tau_{a/r}=e^{-1/(t_{a/r}f_{s})}$ are applied with prefix log-sums, so the same recurrence can be differentiated.
The smoothed gain $G_{s}$ is the output of this one-pole filter, and the waveform is $y[n]=x[n]\,10^{G_{s}[n]/20}$.

\paragraph{Waveshaping.}
Each of the five curves of \Eqref{eq:waveshaper} has its own learned gains and a bias shared across all inputs (Table~\ref{tab:waveshaper-curves}).
A first-order Butterworth tone filter $|H(f)|=1/\sqrt{1+(f/f_{c})^{2}}$ follows in the frequency domain, and the tone cutoff and makeup gain are also predicted.

\paragraph{Reverberation.}
Algorithmic reverberators realize decay through feedback, as parallel combs or feedback delay networks~\citep{valimaki2012fifty}.
Their impulse response is implicit, and unrolling those loops for tails that last seconds is unstable to differentiate.
We instead synthesize a closed-form decaying-noise impulse whose per-band envelopes match the RT60 that a feedback network would produce.
For each of six bands $b$ with $\delta_{b}=3\ln 10\,/\,\mathrm{RT60}_{b}$,
\begin{equation}
  h_{\mathrm{late},b}[n] = \varepsilon_{b}[n]\,e^{-\delta_{b}\,n/f_{s}},
  \label{eq:reverb-decay}
\end{equation}
where $\varepsilon_{b}$ is deterministic band-filtered noise.
The bands are summed, and a broadband damping envelope $(1-\eta)+\eta\,e^{-\kappa n/f_{s}}$ with predicted depth $\eta$ and rate $\kappa$ shortens the whole tail.
This envelope is not frequency-selective, so any frequency-dependent decay comes from the per-band RT60 values.
Sparse early reflections with a predicted density are mixed with the tail by a predicted amplitude ratio $r$, as $(1-r)\,h_{\mathrm{late}}+r\,h_{\mathrm{early}}$, and a unit direct impulse is added at $n=0$.
The head predicts the six per-band RT60 values, the damping depth and rate, the early-reflection density, and the early/late ratio.
The ConvNeXt tier that refines this impulse is pretrained separately and kept frozen while the residual is trained.
We have not evaluated $g$ with the parametric impulse alone.

\section{Residual architecture comparison}
\label{app:nn-sweep}

Twenty-three residual architectures, including feed-forward, recurrent~\citep{cho2014learning}, temporal-convolutional, state-space~\citep{gu2021efficiently,gupta2022diagonal}, and self-attentive blocks, were trained on the same frozen parametric decoder and evaluated on 23 stratified held-out pairs (Table~\ref{tab:nn-sweep-full}).
Rank 1 is highest SI-SDR.
The three highest-ranked architectures all have temporal capacity: the causal TCN used in $g$, a grouped SRU (rank 2), and an S4D hybrid (rank 3), within $0.1$\,dB of one another; the SRU and S4D variants are worse on log-spectral distance.
The best feed-forward block (AdaLN) ranks fourth, about 0.4\,dB below the causal TCN.

\begin{table}[h]
\caption{Residual architecture comparison (23 pairs; the four highest-ranked of 23 architectures).
Temporal means recurrent, convolutional-over-time, or attention-over-time capacity.}
\label{tab:nn-sweep-full}
\begin{center}
\begin{small}
\begin{tabular}{llccccc}
\toprule
\textbf{Rank} & \textbf{Architecture} & \textbf{Temporal} & \textbf{SI-SDR (dB)} & \textbf{LSD} & \textbf{L1} & \textbf{CLAP Dist.} \\
\midrule
1  & Causal TCN (used)                   & Yes & $-19.29$ & 0.871 & 0.184 & 0.405 \\
2  & Grouped SRU                         & Yes & $-19.31$ & 0.894 & 0.183 & 0.366 \\
3  & S4D hybrid                          & Yes & $-19.37$ & 0.968 & 0.181 & 0.392 \\
4  & AdaLN feed-forward                  & No  & $-19.72$ & 0.940 & 0.181 & 0.379 \\
\midrule
\multicolumn{7}{l}{\textit{Ranks 5--23 (19 architectures): SI-SDR from $-20.52$ to $-27.04$\,dB}} \\
\bottomrule
\end{tabular}
\end{small}
\end{center}
\end{table}

\section{Decoder curricula}
\label{app:curricula}

Tables~\ref{tab:joint-separate} and~\ref{tab:mushra-study2} compare the three curricula of Section~\ref{subsec:joint-training}, objectively and by listening (Study~2).
Both use same-content reconstruction on validation clips: $z$ is extracted from each clip's processed signal and applied to its own dry signal, unlike the cross-content pairs of Section~\ref{sec:results}.
Spectral residuals favor joint training from scratch.
In Study~2, every listener rated the two-stage curriculum (joint finetune) above separate heads and above dry audio.
Both contrasts gave $p=0.0312$, the smallest value attainable with six listeners, which does not survive a Holm correction for the two contrasts ($0.0625$).
Joint training from scratch scored about the same as dry audio (48.8 versus 49.7).

\begin{table}[h]
\caption{Joint versus separate parameter-head training (200 validation-split clips, same-content reconstruction).
CLAP Dist.\ is $1-\cos$ between CLAP embeddings of prediction and reference.}
\label{tab:joint-separate}
\begin{center}
\begin{small}
\begin{tabular}{lccc}
\toprule
\textbf{Condition} & \textbf{STFT L1} & \textbf{Time L1} & \textbf{CLAP Dist.} \\
\midrule
Separate & 2.314 & 0.117 & 0.333 \\
Joint finetune & 1.224 & \textbf{0.116} & 0.308 \\
Joint scratch & \textbf{1.052} & 0.132 & \textbf{0.208} \\
\bottomrule
\end{tabular}
\end{small}
\end{center}
\end{table}

\begin{table}[h]
\caption{Listening results for the curricula of the DSP chain on same-content validation clips (Study~2, $N=6$; similarity to the reference, 0--100).
Means and SDs are over all ratings, and 95\% CIs are bootstrap intervals over per-listener means.}
\label{tab:mushra-study2}
\begin{center}
\begin{small}
\begin{tabular}{lccc}
\toprule
\textbf{Condition} & \textbf{Mean} & \textbf{SD} & \textbf{95\% CI} \\
\midrule
Joint finetune   & \textbf{66.4} & 25.6 & [57.4, 74.2] \\
Separate         & 51.1 & 26.1 & [41.4, 60.2] \\
Unprocessed      & 49.7 & 26.0 & [41.1, 57.9] \\
Joint scratch    & 48.8 & 26.5 & [39.0, 58.6] \\
\bottomrule
\end{tabular}
\end{small}
\end{center}
\end{table}

\section{Reverberation tiers}
\label{app:reverb-tiers}

\begin{table}[h]
\caption{Parametric reverberation prior versus the refined impulse (50 clips).
Each clip is a validation-split dry recording convolved with an impulse response from the reverberation plugins; $z$ is extracted from that convolved clip, so this is same-content reconstruction, and the impulse set is not split.
STFT L1 and time L1 compare the predicted impulse response with the true one, not the rendered audio.
Study~4, which rated the rendered clips, detected no significant listening difference.}
\label{tab:reverb-ablation}
\begin{center}
\begin{small}
\begin{tabular}{lccc}
\toprule
\textbf{Condition} & \textbf{STFT L1 (mean)} & \textbf{STFT L1 (median)} & \textbf{Time L1} \\
\midrule
Parametric prior & 4.451 & 2.873 & 0.00198 \\
Refined impulse & 4.927 & 4.158 & 0.00099 \\
\bottomrule
\end{tabular}
\end{small}
\end{center}
\end{table}

On the impulse responses, refinement halves time-domain L1 and worsens STFT L1 (Table~\ref{tab:reverb-ablation}).
In Study~4 ($N=9$), unprocessed dry audio scored $60.5$, the parametric prior $57.8$, and the refined impulse $56.0$, with no pairwise test reaching $p<0.05$.
The study detected no preference between the tiers, or for either tier over dry audio (Section~\ref{subsec:invertibility-results}).
With nine listeners, it cannot rule out a small one.
The reverberation row of Table~\ref{tab:per-category} (time-domain L1 $0.391$) measures rendered audio of cross-content test pairs through the full decoder and is not comparable with this table.

\section{Cross-content results}
\label{app:transfer}

Table~\ref{tab:per-category} splits the cross-content results of Section~\ref{subsec:invertibility-results} by family.
Among families with at least six pairs, dynamics have the lowest STFT L1.
Chain configurations route several processors in series and so require several modules of $g$ to cooperate, and they are close to the global mean.
CLAP distance ranks the families differently: equalization has the highest STFT L1 but a mid-range CLAP distance, and modulation and chains are farthest from the reference in CLAP space.
These 60 cross-content pairs differ from the 200 same-content validation clips of Table~\ref{tab:joint-separate}, so absolute values are not comparable across the two.

\begin{table}[h]
\caption{Per-family cross-content reconstruction after adversarial fine-tuning (60 pairs, R3GAN variant).
Chain is an evaluation grouping for held-out configurations that route several processors in series, not one of the eight plugin families of Table~\ref{tab:plugin-families}.
One pair with an unresolved category is included in the global row only.}
\label{tab:per-category}
\begin{center}
\begin{small}
\begin{tabular}{lcccc}
\toprule
\textbf{Family} & \textbf{STFT L1} & \textbf{Time L1} & \textbf{CLAP Dist.} & \textbf{$N$} \\
\midrule
Preamplification      & 0.713 & 0.057 & 0.028 &  1 \\
Dynamics              & 0.927 & 0.027 & 0.073 & 11 \\
Modulation            & 1.010 & 0.058 & 0.341 &  6 \\
Reverberation         & 1.015 & 0.391 & 0.184 &  3 \\
Saturation/Distortion & 1.149 & 0.228 & 0.133 &  2 \\
Chain                 & 1.176 & 0.057 & 0.308 & 27 \\
Equalization          & 1.299 & 0.060 & 0.113 &  9 \\
\midrule
\textit{Global}       & \textit{1.120} & \textit{0.075} & \textit{0.220} & \textit{60} \\
\bottomrule
\end{tabular}
\end{small}
\end{center}
\end{table}

\section{Effect-score heuristic}
\label{app:effect-score}

The composite score drops renders whose wet signal is not a salient transformation of the dry signal.

If the wet waveform is silent while the dry waveform is not, the score is $s=-1$ and the pair is never used.

Six normalized discrepancies $d_{k}\in[0,1]$ are then aggregated.
With Pearson correlation $\rho$ between aligned frames, $d_{1}=1-|\rho|$.
With $\Delta_{\mathrm{rms}}$ the absolute RMS difference in decibels, $d_{2}=\min(|\Delta_{\mathrm{rms}}|/12,1)$.
Crest-factor, centroid, and spectral-flatness shifts give
$d_{3}=\min(|\Delta_{\mathrm{crest}}|/6,1)$,
$d_{4}=\min(|\Delta_{\mathrm{cnt}}|/5000,1)$,
$d_{6}=\min(|\Delta_{\mathrm{flat}}|/0.1,1)$.
Spectra are split into bands below $300\,\mathrm{Hz}$, $300\,\mathrm{Hz}$ to $4\,\mathrm{kHz}$, and above $4\,\mathrm{kHz}$;
$d_{5}=\min(\Delta_{\mathrm{bands}}/0.5,1)$ from the $\ell_{1}$ change in band energy fractions.
Weights $(1,\tfrac12,1,1,1,1)$ downweight RMS-only differences that loudness normalization removes,
\begin{equation*}
 s=\frac{\sum_{k=1}^{6} w_{k} d_{k}}{\sum_{k=1}^{6} w_{k}}\in[0,1].
\end{equation*}

Corpus generation redraws parameters that score below $0.40$.
After redrawing, about $37\%$ of the roughly two million renders still score below $0.40$, as estimated from a histogram of the scores.
The training configurations set a lower floor per stage: $s\ge 0.15$ for the encoder and the residual, and $s\ge 0.30$ for the joint DSP stages and the adversarial stage.
The saved outputs do not record whether this floor was applied to the 60 evaluation pairs.

\section{Plugin catalog}
\label{app:plugins}

Sixty-eight VST3 processors entered training after their rendered wet signals were checked.
Each plugin carries one of 21 fine-grained category tags (the Cat.\ column), which group into the eight families of Table~\ref{tab:plugin-families}.
The search comparison groups its 60 pairs by the label of their configuration, which is a tag or, for configurations that route several plugins in series, ``chain''.
Table~\ref{tab:plugin-catalog} lists identifiers, display names, category tags, and control counts.

Synthetic sources are sine sweeps and white, pink, and brown noise. Datasets distributed at lower sample rates, such as LibriSpeech at 16\,kHz, are resampled to 48\,kHz before rendering, and every source is rendered at full length.
Each training item requires at least two distinct tracks through the same configuration, and configurations that cannot meet that rule are dropped.
Table~\ref{tab:plugin-families} summarizes the 68 plugins by family.

\begin{table}[h]
\caption{Plugin inventory by family (68 total).}
\label{tab:plugin-families}
\begin{center}
\begin{small}
\begin{tabular}{lcp{5.5cm}}
\toprule
\textbf{Family} & \textbf{Plugins} & \textbf{Representative types} \\
\midrule
Dynamics / Compression & 19 & Feed-forward, vari-mu, optical, bus comp \\
Equalization & 14 & Parametric, graphic, vintage transformer \\
Modulation & 9 & Chorus, phaser, flanger, tremolo \\
Preamplification & 9 & Channel strip, transformer, tube pre \\
Reverberation & 8 & Algorithmic, plate, hall, room \\
Analog tape & 5 & Tape saturation, wow/flutter \\
Saturation / Distortion & 3 & Harmonic exciter, soft-clip, wavefold \\
Miscellaneous & 1 & Multi-tool (EQ and waveshaping) \\
\midrule
\textbf{Total} & 68 & \\
\bottomrule
\end{tabular}
\end{small}
\end{center}
\end{table}

\FloatBarrier

\begin{footnotesize}
\begin{longtable}{@{} l p{0.38\linewidth} p{0.13\linewidth} r @{}}
\caption{Sixty-eight training plugins with stable identifiers, display names, condensed category tags, and differentiable control counts ($P_{\mathrm{eff}}$).}%
\label{tab:plugin-catalog}\\
\toprule
ID & Display name & Cat. & $P_{\mathrm{eff}}$ \\
\midrule
\endfirsthead
\multicolumn{4}{c}{\textbf{Table \ref{tab:plugin-catalog} (continued)}}\\
\toprule
ID & Display name & Cat. & $P_{\mathrm{eff}}$ \\
\midrule
\endhead
\midrule
\multicolumn{4}{r}{\textit{(Continued on next page)}}\\
\endfoot
\bottomrule
\endlastfoot
4k\_eq & 4K EQ & eq & 20 \\
airwindows\_chorus & Airwindows Consolidated (Chorus) & chorus & 10 \\
airwindows\_consolidated\_compressor & Airwindows Consolidated (Compressor) & compressor & 10 \\
airwindows\_consolidated\_eq & Airwindows Consolidated (EQ Algorithms) & eq & 10 \\
airwindows\_consolidated\_modulation & Airwindows Consolidated (Modulation Algorithms) & modulation & 10 \\
airwindows\_consolidated\_preamp & Airwindows Consolidated (Console/Preamp) & preamp & 10 \\
airwindows\_consolidated\_reverb & Airwindows Consolidated (Reverb Algorithms) & reverb & 10 \\
airwindows\_tape & Airwindows Consolidated (Tape Algorithms) & tape & 10 \\
airwindows\_tremolo & Airwindows Consolidated (Tremolo) & tremolo & 10 \\
blendeq & BLENDEQ & eq & 12 \\
blue\_cat\_chorus & Blue Cat's Chorus & chorus & 5 \\
blue\_cat\_flanger & Blue Cat's Flanger & flanger & 7 \\
blue\_cat\_phaser & Blue Cat's Phaser & phaser & 6 \\
blue\_cat\_triple\_eq & Blue Cat Triple EQ & eq & 3 \\
britpre & BritPre & preamp & 4 \\
britpressor & Britpressor & vca compressor & 7 \\
busterse & BUSTERse & compressor & 10 \\
byod & BYOD & distortion & 1 \\
caelum\_tape\_cassette\_2 & Tape Cassette 2 & tape & 7 \\
chow\_centaur & ChowCentaur & overdrive & 4 \\
chow\_multitool & ChowMultiTool & eq waveshaper & 2 \\
chow\_phaser & ChowPhaser & phaser & 7 \\
ctagdrc & CTAGDRC & compressor & 11 \\
dbcomp & dBComp & compressor & 1 \\
dc1a & DC1A & compressor & 6 \\
dragonfly\_early\_reflections & Dragonfly Early Reflections & reverb & 5 \\
dragonfly\_hall\_reverb & Dragonfly Hall Reverb & hall reverb & 17 \\
dragonfly\_plate\_reverb & Dragonfly Plate Reverb & plate reverb & 5 \\
dragonfly\_room\_reverb & Dragonfly Room Reverb & room reverb & 18 \\
fetcb & FetCB & compressor & 4 \\
fetish & FETish & compressor & 9 \\
g395a & G395a & vintage eq & 4 \\
harqules & Harqules & console eq & 10 \\
ivgi & IVGI2 & saturation & 7 \\
kissofshame & The Kiss of Shame & tape & 5 \\
kolin & Kolin & compressor & 5 \\
lala & LALA & compressor & 4 \\
maxbax & MAXBAX & eq & 3 \\
mjuc\_jr & MJUC jr. & compressor & 4 \\
n492me & N492ME (NEQME) & eq & 6 \\
oss & OSS & compressor & 1 \\
prebox & PreBOX & preamp & 4 \\
predd & PREDD & preamp & 8 \\
protoverb & Protoverb & reverb & 2 \\
relief\_compressor & Relief Compressor & compressor & 18 \\
relief\_eq & Relief EQ & eq & 20 \\
specomp & SPECOMP & compressor & 3 \\
ssq & SSQ & eq & 11 \\
t\_saturator & T-Saturator & tape & 11 \\
tal\_chorus\_lx & TAL-Chorus-LX & chorus & 3 \\
tal\_reverb\_4 & TAL-Reverb-4 & plate reverb & 17 \\
tapemachine & TapeMachine & tape & 17 \\
tdr\_kotelnikov & TDR Kotelnikov & compressor & 7 \\
tdr\_nova & TDR Nova & eq & 45 \\
tdr\_vos\_slickeq & TDR VOS SlickEQ & eq & 11 \\
tilta & TILTA & eq & 3 \\
tuba & TUBA & preamp & 7 \\
tube\_preamp & Tube Preamp & preamp & 15 \\
tupre & TuPRE & preamp & 4 \\
valhalla\_space\_modulator & Valhalla Space Modulator & flanger & 5 \\
valhalla\_supermassive & Valhalla Supermassive & reverb & 11 \\
varimoon & VariMoon & compressor & 3 \\
vpre72 & VPRE-72 & preamp & 4 \\
xmcompressor & XMCompressor & compressor & 5 \\
xmeq & XMEQ & eq & 4 \\
xmtubepre & XMTubePre & preamp & 1 \\
yala & YALA & vari mu compressor & 3 \\
zl\_compressor & ZL Compressor & compressor & 17 \\
\end{longtable}
\end{footnotesize}

\section{Architecture details}
\label{app:arch}

Table~\ref{tab:app-encoder} specifies the encoder front end and Conformer stack, Table~\ref{tab:app-actuator} gives per-module actuator ranges, and Table~\ref{tab:waveshaper-curves} lists the waveshaper curves mixed by \Eqref{eq:waveshaper}.

\begin{table}[h]
\begin{center}
\begin{footnotesize}
\caption{Encoder front end and Conformer stack.}
\label{tab:app-encoder}
\begin{tabular}{@{}p{0.22\linewidth} p{0.72\linewidth}@{}}
\toprule
\textbf{Component} & \textbf{Specification (8-layer; 6-layer omits two blocks)} \\
\midrule
Frontend & Hann FFT, $N_{\mathrm{FFT}}=2048$, hop $960$ at $48\,\mathrm{kHz}$; $80$ mel bins from $0$ to $24\,\mathrm{kHz}$, decibel scale with an $80$\,dB floor. \\
Stem & Convolutional downsampler, $8\times$ temporal shrink, width $128$. \\
Conformer & Width $256$, depth $N\in\{8,6\}$, 4 heads, depthwise conv kernel $9$, FFN $\times 4$, rotary/content mix $50$:$50$, DropPath $0.1$. \\
Pooling & Learned query token, then an affine map to $z\in\mathbb{R}^{256}$. \\
Size & About $5$ to $6\times 10^{6}$ parameters for $N=8$; $3$ to $4\times 10^{6}$ for $N=6$. \\
\bottomrule
\end{tabular}
\end{footnotesize}
\end{center}
\end{table}

\begin{table}[h]
\begin{center}
\begin{footnotesize}
\caption{Actuator spans after sigmoidal remapping.}
\label{tab:app-actuator}
\begin{tabular}{ll}
\toprule
\textbf{Stage} & \textbf{Ranges} \\
\midrule
EQ & Gain $\pm 12$\,dB; $Q\in[0.1,10]$; frequency $20\,\mathrm{Hz}$ to $20\,\mathrm{kHz}$ \\
Compressor & Threshold $[-60,0]$\,dBFS; ratio $[1,20]$; attack $[1,100]\,\mathrm{ms}$; release $[10,1000]\,\mathrm{ms}$ \\
Distortion & Drive up to $+36$\,dB; tone low-pass $500\,\mathrm{Hz}$ to $20\,\mathrm{kHz}$; makeup $-12$ to $+6$\,dB \\
\bottomrule
\end{tabular}
\end{footnotesize}
\end{center}
\end{table}

\begin{table}[h]
\begin{center}
\begin{footnotesize}
\caption{Waveshaper curves mixed by the softmax in \Eqref{eq:waveshaper}.}
\label{tab:waveshaper-curves}
\begin{tabular}{@{}lll@{}}
\toprule
$k$ & Name & $s_{k}(x)$ \\
\midrule
1 & Tanh saturation & $g_{o}\tanh(g_{i}x+b)$ \\
2 & Hard clip & $g_{o}\,\mathrm{clip}(g_{i}x+b,-1,1)$ \\
3 & Asymmetric tube & $g_{o}\tanh\!\big(g_{\pm}(x+b)\big)$, $g_{+}$ if $x+b\ge0$, else $g_{-}$ \\
4 & Diode & $g_{o}\,\dfrac{u}{\sqrt{u^{2}+\varepsilon}}\big(1-e^{-\kappa|u|}\big)$, $u=x+b$ \\
5 & Sine wavefolder & $g_{o}\sin(g_{i}x+b)$ \\
\bottomrule
\end{tabular}
\end{footnotesize}
\end{center}
\end{table}

Residual color is a causal TCN: a pointwise input convolution to 32 channels; four blocks of dilated causal convolution (kernel 13, dilation $2^{i}$), batch normalization, PReLU, and FiLM (width 128), each with a residual connection; a pointwise output convolution; and a skip connection from the input.
Module gates are continuous sigmoids, initialized open ($\gamma_{m}\approx0.88$).

\section{Training hyperparameters}
\label{app:hparams}

Unless noted, optimization is AdamW~\citep{loshchilov2019decoupled}, that is, Adam~\citep{kingma2014adam} with decoupled weight decay, with linear warmup and cosine decay.

\textbf{Encoder classification.}
Peak learning rate $3\times 10^{-4}$, weight decay $5\times 10^{-2}$, batch $32$, dropout $0.4$, warmup $1000$ steps, up to $250{,}000$ updates.
AM-Softmax uses scale 15 and margin 0.25; the VICReg variance and covariance terms~\citep{bardes2022vicreg} have weights $\lambda_{\mathrm{var}}=1.0$ and $\lambda_{\mathrm{cov}}=0.04$.
Pairs scoring below about $0.15$ on Appendix~\ref{app:effect-score} are not mined.

\textbf{Joint DSP prediction.}
Learning rate $3\times 10^{-4}$, weight decay $10^{-4}$, warmup $1000$, batch $64$, $50{,}000$ steps.
Spectral reconstruction uses multi-resolution Hann STFT pairs~\citep{yamamoto2020parallel} with $(N_{\mathrm{FFT}},\mathrm{hop})$ of $(512,128)$, $(1024,256)$, and $(2048,512)$, unit weight, and $\lambda_{\mathrm{time}}=0.1$ on waveform $\ell_{1}$.
At each resolution the training loss sums spectral convergence, the $\ell_{1}$ distance between linear magnitudes, and the $\ell_{1}$ distance between log magnitudes; the evaluation metric of Section~\ref{subsec:evaluation} keeps only the log-magnitude term.
This stage has no gate loss.

\textbf{Residual color.}
$200{,}000$ steps at peak learning rate $10^{-5}$, warmup $4000$, weight decay $10^{-4}$, batch $8$, same STFT settings, and a gate loss with $\lambda_{\mathrm{gate}}=0.5$: the mean squared error between the module gates and a binary pattern per plugin category (for example, equalizer and residual on, the rest off, for an equalizer), skipped for categories without a pattern.

\textbf{Impulse-response refinement (pretrained separately, then frozen while the residual is trained).}
Generator and critic peaks $(2\times 10^{-4},\,10^{-4})$, $100{,}000$ iterations, batch 8, adversarial weight $10^{-2}$, feature matching $10^{-1}$.

\textbf{Perceptual refinement of $g$.}
Both variants: generator peak $5\times 10^{-6}$, critic $2.5\times 10^{-5}$, batch 4, $50{,}000$ iterations, reconstruction weights $\lambda_{\mathrm{FFT}}=1$, $\lambda_{\mathrm{wave}}=0.1$, $\lambda_{\mathrm{gate}}=0.5$, adversarial weight $10^{-2}$ ramped in over $10{,}000$ steps after $5{,}000$ reconstruction-only steps, feature matching $1.0$, gradient clipping $0.1$ (generator) and $0.5$ (critic).
One critic step per three generator steps.
R1 and R2 penalties have weight $\gamma=10$, decayed by $0.9999$ per step to a floor of $1$, and are computed every 16 critic steps.
The optimizer covers all system parameters, including the encoder's, unless the encoder is frozen as an ablation.

\section{Search and DeepAFx-ST comparison details}
\label{app:search-family}

\paragraph{Reduced setting.}
The reduced setting uses a population of 8 with $\sigma=0.1$, started from the best of 8 uniformly random candidates, for 128 renders per example with no early stopping.
It matches the five-second clips directly, without the padding of the public code.
At this setting we also searched in TAPE's space, with $f$ and $g$ frozen.
\emph{Latent CMA-ES} perturbs $z$ around $f(\mathrm{wet}_{A})$ (32 candidates, $\sigma=0.2$ relative to the RMS of $z$) and decodes each candidate with $g$.
\emph{Latent TTA} takes 16 Adam steps on $z$ (learning rate $0.01$) against the same two objectives.
The \emph{oracle} takes the same 16 steps against a multi-resolution STFT loss to $\mathrm{wet}_{B}$.
\emph{TAPE, then chain search} ran at both settings.

Table~\ref{tab:search} gives the results.
After 128 renders, the AFx-Rep search had brought its outputs closer to $\mathrm{wet}_{A}$ in AFx-Rep than TAPE's output (cosine distance 0.39 versus 0.66).
The same outputs were farther from $\mathrm{wet}_{B}$ than both TAPE's output and the unprocessed input (STFT L1 3.28 versus 1.12 and 2.34; $p<10^{-10}$ against TAPE).
Optimizing $f$ instead of AFx-Rep, or searching a correction applied after TAPE's output, did not close the gap.
The corrective search moved TAPE's output away from the target as its budget grew (Figure~\ref{fig:search}).
Part of the raw gap is level.
TAPE's outputs have a mean signed RMS error of $+0.8$\,dB against the target (mean absolute error 5.6\,dB), whereas the chain's peak-normalized outputs are on average about 14\,dB louder than the target.
The gap persists after peak-normalizing output and target (2.40 for ST-ITO, 1.39 for TAPE, and 1.81 for the input).
CMA-ES lowered the AFx-Rep distance more than random search did (0.39 versus 0.51; $p<10^{-7}$), but its STFT L1 was not lower (3.28 versus 3.03; $p=0.51$).
With $f$ as the objective, the error fell from 3.14 at this setting to 2.50 at the full setting ($p<10^{-4}$).

\begin{table}[h]
\caption{One forward pass versus per-example search at the reduced setting (60 held-out pairs; means over pairs).
STFT L1 is measured against the withheld $\mathrm{wet}_{B}$; \emph{Peak} normalizes output and target to unit peak first.
$\Delta$ is the change in STFT L1 relative to TAPE, with a 95\% bootstrap interval.
\emph{AFx} and $f$ are cosine distances to $\mathrm{wet}_{A}$ in AFx-Rep and in our encoder; each search lowers the distance it optimizes.
Renders count full evaluations of the basic chain or of $g$ per example; TTA also takes 16 backward passes.
Time is the median wall-clock seconds per five-second clip.
${}^{\dagger}$Searches of the basic chain with $f$ optimize $f$ on peak-normalized audio, which reaches 0.29 and 0.27; the column shows the unnormalized distance.}
\label{tab:search}
\begin{center}
\begin{footnotesize}
\setlength{\tabcolsep}{4pt}
\begin{tabular}{llrrccccc}
\toprule
\textbf{Method} & \textbf{Objective} & \textbf{Renders} & \textbf{Time} & \textbf{STFT L1} & \textbf{$\Delta$ [95\% CI]} & \textbf{Peak} & \textbf{AFx} & \textbf{$f$} \\
\midrule
Unprocessed $\mathrm{dry}_{B}$ & -- & 0 & -- & 2.335 & $+1.215$ [$+0.929$, $+1.537$] & 1.808 & 0.894 & 0.797 \\
TAPE, one pass & -- & 1 & 0.52 & 1.120 & -- & 1.391 & 0.660 & 0.372 \\
\midrule
\multicolumn{9}{l}{\textit{Search over ST-ITO's basic chain}} \\
ST-ITO search & AFx-Rep & 128 & 4.2 & 3.278 & $+2.158$ [$+1.687$, $+2.709$] & 2.395 & 0.394 & 0.673 \\
ST-ITO search & $f$ & 128 & 6.1 & 3.143 & $+2.023$ [$+1.689$, $+2.382$] & 2.255 & 0.819 & 0.510$^{\dagger}$ \\
Random search & AFx-Rep & 128 & 4.1 & 3.033 & $+1.913$ [$+1.545$, $+2.312$] & 2.288 & 0.508 & 0.733 \\
TAPE, then chain search & AFx-Rep & 129 & 4.7 & 2.859 & $+1.739$ [$+1.297$, $+2.207$] & 2.177 & 0.355 & 0.601 \\
TAPE, then chain search & $f$ & 129 & 6.6 & 2.687 & $+1.567$ [$+1.203$, $+1.962$] & 2.040 & 0.677 & 0.458$^{\dagger}$ \\
\midrule
\multicolumn{9}{l}{\textit{Search in TAPE's space, from $f(\mathrm{wet}_{A})$}} \\
Latent CMA-ES & AFx-Rep & 33 & 16.8 & 1.119 & $-0.001$ [$-0.033$, $+0.029$] & 1.368 & 0.587 & 0.358 \\
Latent CMA-ES & $f$ & 33 & 17.2 & 1.124 & $+0.004$ [$-0.026$, $+0.033$] & 1.398 & 0.642 & 0.293 \\
Latent TTA & AFx-Rep & 18 & 18.9 & 1.139 & $+0.018$ [$-0.039$, $+0.086$] & 1.359 & 0.507 & 0.346 \\
Latent TTA & $f$ & 18 & 19.6 & 1.201 & $+0.081$ [$+0.015$, $+0.158$] & 1.465 & 0.631 & 0.234 \\
Oracle TTA & $\mathrm{wet}_{B}$ & 18 & 18.8 & 0.956 & $-0.164$ [$-0.210$, $-0.120$] & 1.281 & 0.621 & 0.384 \\
\bottomrule
\end{tabular}
\end{footnotesize}
\end{center}
\end{table}

Table~\ref{tab:search-family} splits the comparison of Section~\ref{subsec:search} at the reduced setting by effect family, with the grouping of Table~\ref{tab:per-category}.
The basic-chain search has a higher error than TAPE's single pass in every family, under either objective, and Table~\ref{tab:fullbudget-family} shows the same at the full setting.

\paragraph{Full setting.}
The full setting reruns the basic-chain searches on all 60 pairs with the CMA-ES settings of ST-ITO's public evaluation script (Section~\ref{subsec:search}).
Each candidate is padded to 262{,}144 samples, as in that code.
With early stopping disabled, every search spends its 4{,}096 candidate renders, and the reported output is the best candidate under the padded objective rendered again at the clip length; its objective value can therefore differ slightly from the one used for selection.

\paragraph{Precision and timing.}
Times exclude model loading.
Section~\ref{subsec:realtime} compares the streamed chain with TAPE's output with TF32 disabled in convolutions and matrix products.

\paragraph{DeepAFx-ST.}
We use the two autodiff models of the public DeepAFx-ST release, whose renderer is a differentiable parametric equalizer and compressor, not the retrained DeepAFx-ST+ models of ST-ITO.
On both checkpoints, our wrapper reproduced the predicted parameters and output waveforms of the upstream forward pass exactly.
Input and reference are resampled to 24\,kHz and peak-normalized to $-12$\,dBFS, and the output is peak-normalized, as in the official inference script; timing includes this resampling and normalization.
Table~\ref{tab:fullbudget-family} splits the 24\,kHz comparison by family: with the official recipe, the music model has a higher error than the unprocessed input in every family, including dynamics and equalization.

\begin{table}[h]
\caption{Full-setting search and DeepAFx-ST by effect family (60 pairs; lower is better).
Left: STFT L1 at 48\,kHz, with 4{,}097 renders of the basic chain per search.
Right: peak-normalized STFT L1 after resampling output and target to 24\,kHz, for the official recipe and the 48\,kHz rendering ablation of the music model.
One pair with an unresolved category is omitted.}
\label{tab:fullbudget-family}
\begin{center}
\begin{footnotesize}
\setlength{\tabcolsep}{4pt}
\begin{tabular}{lrcccccccc}
\toprule
 & & \multicolumn{4}{c}{\textbf{48\,kHz}} & \multicolumn{4}{c}{\textbf{24\,kHz, peak-normalized}} \\
\cmidrule(lr){3-6}\cmidrule(lr){7-10}
 & & & & \multicolumn{2}{c}{\textbf{ST-ITO search}} & & & \multicolumn{2}{c}{\textbf{DeepAFx-ST, music}} \\
\cmidrule(lr){5-6}\cmidrule(lr){9-10}
\textbf{Family} & \textbf{$N$} & \textbf{Dry} & \textbf{TAPE} & AFx-Rep & $f$ & \textbf{Dry} & \textbf{TAPE} & Official & 48\,kHz \\
\midrule
Dynamics              & 11 & 2.087 & 0.927 & 2.448 & 2.426 & 1.280 & 0.947 & 1.527 & 1.309 \\
Equalization          &  9 & 2.358 & 1.299 & 3.917 & 2.160 & 2.572 & 1.842 & 2.767 & 2.554 \\
Modulation            &  6 & 1.763 & 1.010 & 2.762 & 2.327 & 0.967 & 1.043 & 1.358 & 1.210 \\
Reverberation         &  3 & 2.055 & 1.014 & 2.034 & 1.362 & 1.579 & 0.914 & 2.012 & 1.728 \\
Saturation/Distortion &  2 & 1.547 & 1.149 & 1.616 & 1.829 & 1.618 & 1.312 & 2.010 & 1.836 \\
Preamplification      &  1 & 1.796 & 0.713 & 1.921 & 1.104 & 1.545 & 1.081 & 2.393 & 2.088 \\
Chain                 & 27 & 2.677 & 1.176 & 3.056 & 2.933 & 1.821 & 1.368 & 2.290 & 2.067 \\
\bottomrule
\end{tabular}
\end{footnotesize}
\end{center}
\end{table}

\begin{table}[h]
\caption{STFT L1 by effect family for the comparison of Table~\ref{tab:search} (60 pairs; lower is better).
Searches use 128 renders of the basic chain, 32 latent candidates, or 16 TTA steps.
One pair with an unresolved category is omitted.}
\label{tab:search-family}
\begin{center}
\begin{footnotesize}
\setlength{\tabcolsep}{4pt}
\begin{tabular}{lrccccccc}
\toprule
 & & & & \multicolumn{2}{c}{\textbf{ST-ITO search}} & \textbf{Latent CMA-ES} & \textbf{Latent TTA} & \\
\cmidrule(lr){5-6}
\textbf{Family} & \textbf{$N$} & \textbf{Dry} & \textbf{TAPE} & AFx-Rep & $f$ & AFx-Rep & $f$ & \textbf{Oracle} \\
\midrule
Dynamics              & 11 & 2.087 & 0.927 & 2.997 & 3.403 & 0.948 & 0.927 & 0.743 \\
Equalization          &  9 & 2.358 & 1.299 & 4.030 & 3.141 & 1.390 & 1.412 & 1.047 \\
Modulation            &  6 & 1.763 & 1.010 & 3.037 & 3.000 & 1.014 & 1.242 & 0.877 \\
Reverberation         &  3 & 2.055 & 1.014 & 1.929 & 1.922 & 1.037 & 1.017 & 0.955 \\
Saturation/Distortion &  2 & 1.547 & 1.149 & 1.573 & 2.287 & 1.152 & 0.943 & 0.923 \\
Preamplification      &  1 & 1.796 & 0.713 & 1.660 & 1.778 & 0.670 & 0.759 & 0.576 \\
Chain                 & 27 & 2.677 & 1.176 & 3.579 & 3.309 & 1.132 & 1.282 & 1.035 \\
\bottomrule
\end{tabular}
\end{footnotesize}
\end{center}
\end{table}

\section{Real-time measurements}
\label{app:benchmarks}

Section~\ref{subsec:realtime} uses two C++ programs: one that runs the trained model, and an allocation-free implementation of the same causal chain that we have run only with random weights.
In both, everything that depends only on $z$, namely filter coefficients, FiLM parameters, and the impulse spectrum, is computed once per reference.

\paragraph{Trained-weight program.}
An exporter loads the R3GAN checkpoint of Section~\ref{subsec:search}, writes the trained tensors of the encoder, the control heads, the gate network, the TCN, and the impulse generator, and checks every exported tensor against the checkpoint.
The C++ encoder matches PyTorch at 131\,dB on a random input and at no less than 109.9\,dB on the 60 references.
The streaming mode keeps the state of the five biquads, the compressor's RMS detector and all five passes of its attack--release recursion, the tone filter and DC blocker, each TCN layer's input history together with its FiLM modulation, and a uniformly partitioned FFT convolution of the learned impulse.
It runs on one CPU core, with timing not isolated from other workloads, and simulates host callbacks offline.
It still allocates memory inside callbacks.
Conditioning (encoding the reference, predicting controls, and generating the impulse) takes 1.43\,s per reference on average and is excluded from callback times.
Besides the streaming mode, the program has a whole-clip mode with the offline operators, used for the parity check in Section~\ref{subsec:realtime}, and a reset mode that processes each buffer from a cleared state.
In whole-clip mode, its mean STFT L1 against the target differs from that of PyTorch's strict single-precision output by $3\times10^{-5}$, so its arithmetic differences are much smaller than those that streaming introduces.
Across clips, streaming changed STFT L1 by $-0.11$ to $+0.35$, and both the streamed and the strict offline outputs lie about 0.06 above TAPE's GPU output with TF32 (Table~\ref{tab:fullbudget}).
The buffer-size comparison of Section~\ref{subsec:realtime} covers all 60 clips at all five buffer sizes, and the causality check uses 256-sample buffers and fixed conditioning.

\paragraph{Allocation-free implementation.}
On one core of an older server CPU than the one used in Section~\ref{subsec:realtime}, the allocation-free chain processes 64-sample buffers 2.6 times faster than real time on average and 2.3 times at the 99.9th percentile of per-buffer time, with strict floating-point arithmetic.
These values are medians over five 60\,s runs.
In five separate 10\,s profiling runs, the TCN took about three quarters of the time and the reverberator about a fifth.
The two programs differ in code and processor, so their timings are not directly comparable.
The allocation-free program is written in C++, runs single-threaded, and makes no memory allocation while processing.
Filter, envelope, and level-detector state is held in double precision, and FFT products and neural layers in single precision.
The compressor streams exactly because its block formulation in training computes the same attack--release recursion, refined in five passes, as a sample-by-sample update (signal-to-error ratio 105\,dB in single precision over 100 random settings).
Each TCN layer keeps its own input history (at most 96 samples for the largest dilation), together spanning the network's 181-sample receptive field, with batch normalization folded into the convolutions.
The reverberator holds a frequency-domain delay line over all partitions of the 144{,}000-sample impulse and returns the valid half of each $2B$-point transform for the current buffer, so no tail is discarded at buffer boundaries.
This program omits the learned impulse generator and the control heads, and a random decaying impulse of the trained length stands in for the generator.

Validation covers four tests.
Processing 10\,s of noise and of a synthetic harmonic signal in buffers of 32 to 1{,}024 samples differs from a single whole-signal call by at most $9.4\times10^{-7}$.
An independent offline implementation of the same operators, run on the same weights and controls with irregular host-call lengths, agrees at a signal-to-error ratio of 119.5\,dB with fast floating-point math and 121.5\,dB without it.
With network biases set to zero, an impulse produces exactly zero output before its onset and a nonzero response at it.
Table~\ref{tab:streaming-gap} reports the gap between each streamed stage and the corresponding operation of the offline model over 100 random settings.

\begin{table}[h]
\caption{Signal-to-error ratio (dB) between streamed stages and the offline formulation used in training, over 100 random settings on 5\,s clips.}
\label{tab:streaming-gap}
\begin{center}
\begin{small}
\begin{tabular}{lcc}
\toprule
\textbf{Stage} & \textbf{Median} & \textbf{5th percentile} \\
\midrule
Equalizer (biquad cascade vs.\ FFT-grid response) & 54.4 & 29.7 \\
Compressor (single precision) & 105.2 & 82.3 \\
Tone filter (one-pole vs.\ zero-phase) & 8.2 & 7.0 \\
Mean removal (DC blocker vs.\ clip mean) & 19.7 & 15.2 \\
Waveshaper stage, all changes & 12.6 & 6.1 \\
\bottomrule
\end{tabular}
\end{small}
\end{center}
\end{table}

\end{document}